\documentclass{iau}
\usepackage{graphicx}

\title[IAUS295.~~ Evolution of the central regions of galaxy clusters since z$\sim$1] 
{Evolution in cluster cores since z$\sim$1}

\author[C. Burke et al.]   
{Claire Burke$^1$, Chris Collins$^1$, John Stott$^2$ \and Matt Hilton$^3$ }

\affiliation{$^1$ Astrophysics Research Institute, Liverpool John Moores University, UK.   \\email: {\tt cb@astro.livjm.ac.uk} \\[\affilskip]
$^2$Extragalactic \& Cosmology Group, Durham University, UK.\\ 
$^3$Astrophysics \& Cosmology Research Unit, University of KwaZulu-Natal, South Africa.}

\pubyear{2013}
\volume{295}  
\pagerange{xx--xx}
\jname{The intriguing life of massive galaxies}
\editors{D. Thomas, A. Pasquali \& I. Ferreras, eds.}
\begin{document}

\maketitle

\begin{abstract}
A large fraction of the stellar mass in galaxy clusters is thought to be contained in the diffuse low surface brightness intracluster light (ICL). Being bound to the gravitational potential of the cluster rather than any individual galaxy, the ICL contains much information about the evolution of its host cluster and the interactions between the galaxies within. However due its low surface brightness it is notoriously difficult to study. We present the first detection and measurement of the flux contained in the ICL at z$\sim$1. We find that the fraction of the total cluster light contained in the ICL may have increased by factors of 2--4 since z$\sim$1, in contrast to recent findings for the lack of mass and scale size evolution found for brightest cluster galaxies. Our results suggest that late time buildup in cluster cores may occur more through stripping than merging and we discuss the implications of our results for hierarchical simulations.
\end{abstract}

The central regions of galaxy clusters are usually dominated by a massive brightest cluster galaxy (BCG), which generally sits at the centre of mass of the cluster; and a diffuse, extended halo of intracluster light (ICL). The study of the ICL can reveal details of the evolution histories and processes occurring within galaxy clusters. A significant fraction of the total stellar mass and luminosity of galaxy clusters is thought to be contained within the ICL, however since it has very low surface brightness it is often difficult to detect. 
Using a sample of 6 X-ray selected clusters at 0.8$\le$z$\le$1.22 with deep J-band imaging from HAWK-I on the VLT we have performed the first detection and measurement of the ICL at high redshift. We detect the ICL down to a surface brightness of $\mu_J \sim$23 mag/arcsec$^2$ and, using a cautious but robust method, find it to contain 1--4\% of the total cluster light. When compared to nearby clusters, we find the ICL to have grown by 2--4 times as a fraction of the total cluster light since z=1.

We also examine the surface brightness profiles of the BCGs in our sample. We find that up to 50\% of the light from the BCG is outside of a 50 kpc radius, suggesting that when measuring photospheric light from BCGs a significant fraction will be contaminated by the ICL, therefore BCG masses measured observationally will include some contribution from the ICL. Recent hierarchical simulations (e.g., De Lucia and Blaizot, 2007, MNRAS, 375, 2) under-predict the masses of BCGs at high redshift by as much as 50\%, therefore we suggest that the inclusion of the ICL in the mass budgets of simulated BCGs may ease this discrepancy.

When considering recent results for the lack of scale size evolution of BCGs (in contrast to the size growth seen for field ellipticals (see Stott et al., 2011, MNRAS, 414,1)), we suggest that evolution and interactions in the central regions of clusters may involve stripping rather than merging causing the ICL to be built up while the BCG remains relatively unchanged.

For full details of our study see Burke et al., 2012, MNRAS, 425, 2058.


\end{document}